\documentstyle[12pt]{article}

\begin{document}
\begin{center}
{\Large \bf
Neutrino flux in the rotating reference frame }
\bigskip

{\large D.L.~Khokhlov}
\smallskip

{\it Sumy State University, R.-Korsakov St. 2, \\
Sumy 40007, Ukraine\\
E-mail: khokhlov@cafe.sumy.ua}
\end{center}

\begin{abstract}
It is considered neutrino flux in the rotating reference frame. 
Due to the rotation of the frame, neutrino is observed as a  
superposition of two states P-transformed one from another. 
Since P-transformation is forbidden for neutrino, 
in the rotating reference frame 
one can detect a half of neutrino flux. 
Due to the rotation of the earth, the detector of neutrinos
can measure a half of the solar neutrino flux 
predicted by the SSM that
may provide a solution for the solar neutrino puzzle.

\end{abstract}

Weak interactions violate P-invariance and C-invariance but
conserve CP-invariance~\cite{Co}. The operations of C-transformation
and of P-transformation are forbidden for neutrinos, but
the operation of CP-transformation is allowed for neutrinos.

Take the rotating reference frame. Turnover of the frame at the
angle $\pi$ corresponds to P-transformation of the frame.
Under rotation, the positions of the frame in the range from $\pi$
to $2\pi$ are P-transformed from the positions of the frame
in the range from $0$ to $\pi$.
Neutrino in the rotating frame is observed in the superpositional
state
\begin{equation}
|\psi>=\frac{1}{\sqrt{2}}(|\nu>+{\rm P}|\nu>).
\label{eq:psi}
\end{equation}
However P-transformation is forbidden for neutrino
thereforer it is forbidden to detect neutrino in the state P$|\nu>$.
Hence when measuring neutrino flux in the rotating frame it is
forbidden to detect a half of neutrino flux.

As known~\cite{Bah}, \cite{Ha} experiments measured solar
neutrino fluxes significantly smaller than those predicted by
the standard solar model (SSM), e.g.~\cite{Tu}, \cite{BP}.
Discrepancy between the theoretical and
experimental neutrino flux from the sun presents the solar
neutrino puzzle.

The Homestake experiment measured~\cite{Cl}
an average $^{37}$Ar production rate by solar neutrinos in $^{37}$Cl of
\begin{equation}
{\rm Rate~(Chlorine) =2.55\pm 0.17(stat)\pm 0.18(syst)}~SNU
\label{eq:rex}
\end{equation}
which is $32\pm 5\%$ of the $8.1^{+1.0}_{-1.2}~SNU$
predicted by the SSM~\cite{BP}.

The Kamiokande II and III experiment reported~\cite{Su}
the data consistent with $^8$B solar neutrino flux of
\begin{equation}
\phi_{\nu_\odot}=[2.9\pm 0.2(stat)\pm 0.3(syst)]
\times 10^6 cm^{-2}s^{-1}
\end{equation}
which is $51\%\pm 9\%$ of that predicted by the SSM~\cite{BP}.

The GALLEX experiment measured~\cite{An} a $^{71}$Ge production
rate by solar neutrinos in $^{71}$Ga of
\begin{equation}
{\rm Rate~(Gallium)=79\pm 10(stat)\pm 6(syst)}~SNU
\end{equation}
which is $60\%\pm 10\%$ of the
$131.5^{+7}_{-6}~SNU$ predicted by the SSM~\cite{BP}.

The SAGE experiment measured~\cite{Ab} an
average $^{71}$Ge production rate by solar neutrinos in $^{71}$Ga of
\begin{equation}
{\rm Rate~(Gallium)=74^{+13}_{-12}(stat)^{+5}_{-7}(sys)}~SNU
\end{equation}
which is $56\%\pm 11\%$ of the $131.5^{+7}_{-6}~SNU$
predicted by the SSM~\cite{BP}.

Due to the rotation of the earth, the detector of neutrinos
can measure a half of the solar neutrino flux 
predicted by the SSM. This is consistent with the
data of the Kamiokande, GALLEX, SAGE experiments and significantly
reduces the discrepancy between the theoretical and
experimental data for the Homestake experiment.
So the effect under consideration
may provide a solution for the solar neutrino puzzle.

\end{document}